\documentclass[journal, 10pt]{IEEEtran}
\usepackage[utf8]{inputenc}
\usepackage[T1]{fontenc}
\usepackage{amsmath, mathrsfs, amssymb, ntheorem, mathtools, breqn}
\usepackage{graphicx}
\DeclareGraphicsExtensions{.png, .jpeg, .pdf}
\graphicspath{{./}{figures/}}
\usepackage{dblfloatfix}
\usepackage[ruled,vlined,linesnumbered]{algorithm2e}
\usepackage{multicol, array, booktabs, siunitx, enumitem}
\usepackage{url}
\usepackage{framed}
\usepackage[caption=false,font=footnotesize]{subfig}
\usepackage{microtype}
\usepackage[dvipsnames]{xcolor}

\def\subparagraph{\paragraph}
 \usepackage[small,compact]{titlesec}

\titlespacing*{\section}{0pt}{*1}{*1}
\titlespacing*{\subsection}{0pt}{*1}{*0}
\titlespacing*{\subsubsection}{0pt}{*1}{*0}
\DeclareMathSizes{8}{8}{8}{8}
\newcommand{\MAT}[1]{\ensuremath{\mathbf{#1}}} 
\newcommand{\SET}[1]{\ensuremath{{#1}}}               
\newcommand{\SEQ}[1]{\ensuremath{\mathcal{#1}}}               

\def\P{\text{\rm P}}
\def\PHAT{\widehat{\P}}

\DeclareMathOperator*{\argmax}{arg\,max\,\, }

\makeatletter
\def\namedlabel#1#2{\begingroup
   \def\@currentlabel{#2}%
   \label{#1}\endgroup
}
\makeatother

\newtheorem{theorem}{Theorem}[section]

\newtheorem{observation}[theorem]{Observation}

\newtheorem{definition}{Definition}

\theoremstyle{definition}

\newtheorem{assumption}{Assumption}

\def\UU{\ensuremath{\mathscr{U}}}
\def\S{\ensuremath{\mathscr{S}}}
\def\PP{\ensuremath{\mathscr{P}}}

\newcommand{\DICT}[1]{\ensuremath{D^{#1}}}
\def\DENY{{\text{(\ensuremath{\epsilon, m})--\-Plaus\-ible Deni\-abil\-ity}}}
\newcommand{\PDD}{\ensuremath{\Delta_{att,k}^{u,c}}}
\newcommand{\PDDPROXY}{\ensuremath{\Delta_{p,k}^{u,c}}}
\newcommand{\THA}{\ensuremath{\Theta_{u,k}^{u,c}}}
\newcommand{\THABASE}{\ensuremath{\Theta^{c}}}

\begin{document}
\title{3PS - Online Privacy through Group Identities}
\author{P\'{o}l Mac Aonghusa and Douglas J. Leith
\thanks{P. Mac Aonghusa is with IBM Research and Trinity College Dublin.}%
\thanks{D.J. Leith is with Trinity College Dublin.}
}%
\markboth{IEEE Transactions on Information Forensics and Security}{Mac Aonghusa and Leith: 3PS - Online Privacy through Group Identities}%
\maketitle

\begin{abstract}
Limiting online data collection to the minimum required for specific purposes is mandated by modern privacy legislation such as the General Data Protection Regulation (GDPR) and the California Consumer Protection Act. 
This is particularly true in online services where broad collection of personal information represents an obvious concern for privacy. 
We challenge the view that broad personal data collection is required to provide personalised services. 
By first developing formal models of privacy and utility, we show how users can obtain personalised content, while retaining an ability to plausibly deny their interests in topics they regard as sensitive using a system of proxy, group identities we call 3PS.    
Through extensive experiment on a prototype implementation, using openly accessible data sources, we show that 3PS provides personalised content to individual users over $98\%$ of the time in our tests, while protecting plausible deniability effectively in the face of worst-case threats from a variety of attack types. 
\end{abstract}
\section{Introduction}
Gathering and analysing data about user interests and behaviours is arguably the de facto business model for the free-to-use internet. Personalisation to enhance user experience is offered as a general motivation for broad data collection. The numbers are impressive. Facebook earned an average of US\$4.65 per user from personalised content such as advertising and promoted posts in the second quarter of 2017, according to the Economist~\cite{economist2017}. By comparison, an average of just US\$0.08 per user came from direct fees such as payments within virtual games.  

In this paper we ask a natural question, is it true that much less personal information than is currently collected is sufficient to provide an effective personalised service?  Recent legislation, such as the General Data Protection Regulation (GDPR), mandates that \emph{personal data must be adequate, relevant and limited to what is necessary in relation to the purposes for which those data are processed}, \cite{GDPR:2016}. In this respect, broad collection of user data without transparent purpose in online interactions with everyday commercial systems is a particular concern for individual privacy.

We consider users of everyday online commercial systems where personalised content, tuned to user interests, is displayed during interactions. The privacy model considered here is based on plausible deniability of likely interest in topics an individual user regards as sensitive. We show that, by adopting the persona of an appropriate group containing many users, an individual user can gain a good degree of personalisation while successfully limiting personal data disclosure. The use of group identities as a proxy technique provides a natural ``hiding in the crowd'' form of privacy comparable to techniques such as k-anonymity, so that a user can plausibly deny their interests in topics they deem sensitive. This model is intuitive for users to understand and, importantly, to appreciate its limitations.

Our contributions include a novel \emph{proxy agent} framework we call \emph{3PS} for \textbf{P}rivacy \textbf{P}reserving \textbf{P}roxy \textbf{S}ervice,  where a user may protect their interests in sensitive topics from unwanted personalisation by submitting queries though a pool of group identities called \emph{Proxy Agents}. We also formalise notions of personalisation utility and privacy detection and test these experimentally using openly available data-sets. We show that user privacy need not come at the cost of reduced utility in personalised services when aggregated group information represented by the proxy agent pool is sufficient for personalisation.  

The 3PS framework is designed to be simple to deploy with minimal technical disruption to existing systems. We provide a privacy preserving algorithm for selecting group membership of proxy agents. By running the selection algorithm locally, users can find the group identity best matching their interests without revealing their interests. Through extensive experimental verification we show that our method of selecting group membership is both accurate - selecting the group identifier closest in topical interests with $98\%$ average accuracy across all experiments - and converges rapidly within $3$ input--output iterations on average. 

Personal privacy is fundamentally a risk management exercise where there is an ongoing responsibility on users to take reasonable care. There are no absolute guarantees and individuals must strike their own balance between privacy and utility. Our results suggest that using group identities such as 3PS can provide effective and verifiable privacy protection for responsible users without overly degrading the personalisation capability of the underlying backend system.
\section{Privacy and Personalisation Models}
\label{sec:ptm}
\subsection{General Setup}
\label{sec:io}
We consider a setup where users interact with a system \S{} by by submitting an  input and receiving an output in response. Each interaction between a user and \S{} consists of  an input--output pair, referred to as an \emph{input--output interaction} or \emph{step}. 
We assume that user inputs and system outputs are each decomposable into \emph{features}. 
For example, when modelling a user querying movies or hotels the input features might consist of keywords, or if assigning ratings the features might be clicks.
An ordered list of features with no duplicate entries is called a \emph{dictionary}. We let $\DICT{X}$ and $\DICT{Y}$ denote the dictionary containing valid input features to \S{}, and valid output features generated by \S{} respectively.  Individual features are indicated thus, $\theta^{X}_{i}, \, i=1,\ldots\vert\DICT{X}\vert$ and $\theta^{Y}_j, \, j=1,\ldots\vert\DICT{Y}\vert$ so that $\theta^{X}_{i}$ denotes the $i^{th}$ feature in $\DICT{X}$ and $\theta^{Y}_j$ the $j^{th}$ feature in \DICT{Y}. 
We let \SET{X} and \SET{Y} denote the sets of possible valid inputs and outputs comprised of combinations of features from $\DICT{X}$ and $\DICT{Y}$ respectively, and the set of valid input--output interactions is $\SET{Z} := \SET{X} \times \SET{Y}$. 

We gather a sequence of consecutive input-output interactions between a user and \S{} into a \emph{session}. Input--output interactions may repeat during a session and so sessions are represented as \emph{sequences} of input--output interactions. We use set notation to improve readability when working with sequences when the meaning as applied to sequences is clear. 
We denote the overall sequence of input--output interactions generated by users up to step $k$ by
 $\SEQ{Z}_{k}:=(z_1, \dots z_k)\cup \SEQ{Z}_{0}$, where $z_j$ denotes the $j^{th}$ element of $\SEQ{Z}_{k}$ and $\SEQ{Z}_{0}$ is the background knowledge available before the first interaction $z_1$ is observed. The subsequence of input--output interactions associated with user $u\in\UU{}$ up to step $k$ is denoted by $\SEQ{Z}_{u,k}:=(z_{u,1}, \dots z_{u,k})\cup \SEQ{Z}_{u,0}$ where $z_{u,j}$ denotes the $j^{th}$ element of $\SEQ{Z}_{u,k}$ and $\SEQ{Z}_{u,0}$ is the background knowledge available about $u$ before $z_{u,1}$ is observed.  

Each user $u$ has a private \emph{labelling function} $l_u:\SET{Z} \longrightarrow \SET{C}$ which associates input--output interactions in $\SET{Z}$ with topic labels selected from a private, user-defined set of labels $\SET{C} = \{c_0, c_1, \ldots c_K \}$.  We adopt the convention that the label $c_0$ is identified with a catch-all ``non-sensitive'' category while the remaining elements in $\SET{C}\setminus\{c_0\} $ label individual ``sensitive'' topics such as ``health'' or ``finances''. The user labelling function $l_u$ is private and labels every input--output pair in $\SEQ{Z}_{u,k}$ with at least one topic from $\SET{C}$.    Mostly we are simply interested in whether an input-output pair is sensitive or not for a user, in that case we define $l_u:\SET{Z} \longrightarrow \{0,1\}$ to be the indicator function with $l_u(z)=1$ when input-output pair $l_u(z)=c$, $c\in\SET{C}\setminus\{c_0\}$, i.e. is labelled with a sensitive topic by user $u$, and $l_u(z)=0$ otherwise.

Let $\SEQ{Z}^{u, c}_{u,k}:= (z\in\SEQ{Z}_{u,k}: l_u(z) = c)$ denote the subsequence of observations originating from user $u$ that are labelled with topic $c\in\SET{C}$. The sequence $\SEQ{Z}^{u, c}_{u,k}:= \bigcup_{c\in\SET{C}\setminus\{c_0\}}\SEQ{Z}^{u, c}_{u,k}$ is the subsequence of observations in $\SEQ{Z}_{u,k}$ that user $u$ has labelled as sensitive. 
We let $\SEQ{Z}_{k}^{u, c}:= (z\in\SEQ{Z}_{k}(k): l_u(z) = c)$  denote the subsequence of $\SEQ{Z}_{k}$ that $u$ would label with topic $c\in\SET{C}$. The sequence $\SEQ{Z}^{u, c}_{k}:= \bigcup_{c\in\SET{C}\setminus\{c_0\}}\SEQ{Z}^{u, c}_{k}$ is the subsequence of $\SEQ{Z}_{k}$ that would be labelled as sensitive by user $u$. The sequence $\SEQ{Z}^{u, c}_{k}$ contains items from users other than $u$.  Consequently, while $\SEQ{Z}^{u, c}_{u, t} \subseteq \SEQ{Z}^{u, c}_{k}$, it is not generally the case that $\SEQ{Z}^{u, c}_{k}$ is a subsequence of $\SEQ{Z}_{u,k}$. 

We assume that user labelling functions are well-behaved in the following sense:
\begin{assumption}[Meaningful Labelling]
\label{assum:consistency}
An input-output pair which is labelled as non-sensitive by a user is truly non-sensitive for that user e.g. the user would be content for it to be shared publicly.   
\end{assumption}%
Assumption \ref{assum:consistency} requires users to strike their own balance between utility and privacy. The low risk strategy of simply labelling every input-output pair as sensitive implies that the user may not be able to use the system at all. For example, if the system is a dating service, the knowledge that a person uses the system necessarily reveals their interest in such a service. A user choosing to use the system cannot include such system-level topics in their sensitive set. The implicit statement in Assumption \ref{assum:consistency} is that users form an individual judgement regarding the inference capabilities of observers and to accept a degree of risk associated with this judgement call proving incorrect. 
\subsection{Privacy and Threat Model}
\label{sec:privacy:model}
Our interest is in privacy attacks where an attacker seeks to infer topics of likely interest to users of online systems. An attacker is successful when users are unable to deny their interest in a topic on the balance of probabilities.
Here attackers have access to input--output interactions $\SEQ{Z}_{att, k}\subseteq\SEQ{Z}_{k}$. By analysing $\SEQ{Z}_{att, k}$ the attacker attempts to estimate topics that are of likely interest to $u$.  
The privacy model here is \emph{plausible deniability}, allowing users to reasonably deny that observations are solely associated with topics they deem sensitive. We formalise plausible deniability in our context as follows: 
\begin{definition}[$\delta$-Plausible Deniability]
\label{def:plausible:deniability}
A user $u$ can plausibly deny their input--output observations are associated with topics they deem sensitive if
\footnote{In this case $\P(z\in\SEQ{Z}^{u, c}_{k} \vert z\in\SEQ{Z}_{att, k})$ denotes $\P(\exists m: l_u(\SEQ{Z}_{att, k}(k)) = 1, m\in\{1,2,\dots\})$.} 
\begin{align}
	\P(z\in\SEQ{Z}^{u, c}_{k} \vert z\in\SEQ{Z}_{att, k}) \leq \delta
	\label{eqn:define:plausible:deniability:0}
\end{align}%
where the deniability parameter, $\delta$, is chosen by $u$ and $\SEQ{Z}_{att, k}$ is the background knowledge of an attacker at step $k$ of a session.
\end{definition}%
This differs from the \DENY{} model introduced in \cite{mac.P2} where an individual user claimed plausible deniability because an input--output observation from that user could be associated with any of several topics.

Observe that
\begin{align}
&\P(z\in \SEQ{Z}^{u, c}_{k} \vert z\in \SEQ{Z}_{att, k}) \notag \\
& \stackrel{(a)}{\leq} \frac{\P(z\in\SEQ{Z}^{u, c}_{k}\cap\SEQ{Z}_{k})}{\P(z\in\SEQ{Z}_{k})}\frac{\P(z\in\SEQ{Z}_{k})}{\P(z\in\SEQ{Z}_{att, k})}
\label{eqn:plausible:deniability:4} \\
& \stackrel{(b)}{=} \frac{\P(z\in\SEQ{Z}^{u, c}_{k}\vert z\in\SEQ{Z}_{k})}{\P(z\in\SEQ{Z}_{att, k} \vert z\in\SEQ{Z}_{k})}
\label{eqn:plausible:deniability:5} 
\end{align}%
where inequality $(a)$ follows from the facts that $\P(z\in \SEQ{Z}^{u, c}_{k} \vert z\in \SEQ{Z}_{att, k})=\P(z\in\SEQ{Z}^{u, c}_{k}\cap\SEQ{Z}_{att, k})/\P(z\in\SEQ{Z}_{att, k})$ and  $\SEQ{Z}^{att}_{u,k}\subseteq\SEQ{Z}_{k}$, and equality $(b)$ follows since $\SEQ{Z}_{att, k}\subseteq\SEQ{Z}_{k}$.  Hence, for $\delta$-plausible deniability to hold it is sufficient that 
\begin{align}
\P(z\in\SEQ{Z}^{u, c}_{k}\vert z\in\SEQ{Z}_{k}) \leq \delta \P(z\in\SEQ{Z}_{att, k} \vert z\in\SEQ{Z}_{k})
\label{eqn:plausible:deniability:5a}	
\end{align}%
From \eqref{eqn:plausible:deniability:5a}, when an observer has access to all of the observations in the system so that $\SEQ{Z}_{att, k} = \SEQ{Z}_{k}$ and $\P(z\in\SEQ{Z}_{att, k} \vert z\in\SEQ{Z}_{k}) = 1$ then it is sufficient to have $\P(z\in\SEQ{Z}^{u, c}_{k}\vert z\in\SEQ{Z}_{k}) \leq \delta$ for $\delta$-plausible deniability to hold.  In the case that the observer is able to make observations at a more local level, so that  $\P(z\in\SEQ{Z}_{att, k} \vert z\in\SEQ{Z}_{k}) = \pi <1$, then \eqref{eqn:plausible:deniability:5a} implies that $\P(z\in\SEQ{Z}^{u, c}_{k}\vert z\in\SEQ{Z}_{k}) \leq \delta \pi $ is required for $\delta$-plausible deniability to hold. Consequently, unless the user can plausibly deny that they contributed to $\SEQ{Z}_{att,k}$, we have 
\begin{observation}[Power of Observers]
\label{prop:1}
Observers represent more powerful threats when they have access to more localised sequences of input--output interactions so there is some trade-off involved in locality versus deniability.
\end{observation}

\subsection{Comparison with Other Privacy Models}
\label{sec:3ps:comarisons}
In the group identity setup considered here, the intention is to deny interest by hiding sensitive user activity in the overall activity of users of shared group identifiers.   
	The setup here can  be compared with other privacy models. We show briefly how this is done in the cases of two common models of privacy, Differential Privacy, \cite{dwork2006differential}, and Individual Re-identification, \cite{sweeney2000simple}. 
\subsubsection{Re-identification}
Re-identification risk occurs when an attacker, possessing observations $\SEQ{Z}_{att, k}$, can assert that sensitive input--output interactions generated by user $u$ are identified with probability greater than $1-\epsilon$ for $0< \epsilon \ll 1$. In other words, when
\begin{align}
\P(z\in\SEQ{Z}^{u, c}_{k} \cap \SEQ{Z}_{u,k} \vert 	z\in\SEQ{Z}_{att, k}) > 1-\epsilon
\label{eqn:re:identification:1}
\end{align}%
for $0<\epsilon \ll 1$.  

If $\delta$-plausible deniability holds \eqref{eqn:define:plausible:deniability:0} guarantees 
\begin{align}
 \P(z\in\SEQ{Z}^{u, c}_{k} \cap \SEQ{Z}_{u,k} \vert 	z\in\SEQ{Z}_{att, k}) \leq \delta
\label{eqn:re:identification:2}	
\end{align}%
since  $\SEQ{Z}^{u, c}_{k} \cap \SEQ{Z}_{u,k}\subseteq\SEQ{Z}^{u, c}_{k}$. 
Consequently \eqref{eqn:define:plausible:deniability:0} prevents re-identification of those sensitive input--output interactions with probability at least $1-\delta$.

\subsubsection{Differential Privacy}
Recall that a query mechanism $\mathbf{M}:\SEQ{D} \longrightarrow \SET{R}$ satisfies $(\epsilon,\gamma)$-differential privacy \cite{dwork2006differential} if, for any two sequences $\SEQ{D}_1,\SEQ{D}_2 \in \SEQ{D}$ of length $n$ differing in one element, and any set of output values $\SET{S} \subseteq  \SET{R}$, we have
\begin{align}
\P(\mathbf{M}(\SEQ{D}_1) \in  \SET{S})	\leq e^{\epsilon}\P(\mathbf{M}(\SEQ{D}_2) \in \SET{S}) + \gamma\label{eq:DP}
\end{align}
One important class of mechanisms are those where sequences in $\SEQ{D}$ are first perturbed, e.g. by adding noise, and then queries are answered.   It is this approach which is effectively adopted here, with the perturbations being introduced by the randomness of the process generating the input--output interactions.
An attacker observes a sequence of  input--output interactions and seeks to associate a label with one or more input--output interactions, namely whether or not they were likely to be generated by a target user $u$ and are sensitive for that user.    Consider therefore the query $\mathbf{M}_{z}(\SEQ{Z}_{k})=l_u(z)$ i.e. which labels input-output pair $z$ as $1$ when it is sensitive for user $u$ and labels it $0$ otherwise.  This is a worst case query in the sense that it assumes the attacker knows the labelling function $l_u$, and when this is not the case the labelling accuracy will obviously be degraded.  Let $\SEQ{D}_1,\SEQ{D}_2 \in \SEQ{D}$ be two input-output sequences such that $\SEQ{D}_1(k)=\SEQ{D}_2(k)$, $k=\{1,\dots,n\}\setminus\{j\}$ where $\SEQ{D}_1(k)$ denotes the $k$'th element of sequence $\SEQ{D}_1$ and similarly for $\SEQ{D}_2(k)$ i.e. sequences  $\SEQ{D}_1$ and $\SEQ{D}_2$ are identical except for the $j$'th element.   Mechanism $\mathbf{M}_{z}$ is $(\epsilon,\gamma)$-differentially private provided
\begin{align}
p_1 \le e^\epsilon p_2 + \gamma&,\ p_2 \le e^\epsilon p_1 + \gamma \label{eq:DP1}\\
1-p_1 \le e^\epsilon (1-p_2) + \gamma&,\ 1-p_2 \le e^\epsilon (1-p_1) + \gamma \label{eq:DP2}
\end{align}
where
\begin{align}
p_1:=\P(l_u(\SEQ{D}_1(j))=1),\ p_2:=\P(l_u(\SEQ{D}_2(j))=1)
\end{align}
are the probabilities that input-output pair $j$ in sequence $\SEQ{D}_1$, respectively $\SEQ{D}_2$, is labelled sensitive by user $u$.  For sequences satisfying the $\delta$-plausible deniability condition (\ref{eqn:define:plausible:deniability:0}) we have $p_1\le \delta$ and $p_2\le \delta$.  It can be verified that the $(\epsilon,\gamma)$-differential privacy conditions (\ref{eq:DP1})-(\ref{eq:DP2}) are therefore satisfied for $\epsilon\ge 0$ and $\gamma\ge\max\{\delta,1-e^\epsilon(1-\delta)\}$.

\subsection{Other Linking Attacks}
{The privacy model described here is concerned with attacks at the application layer that seek to link input--output interactions and associated topics to individual user interests.  Linking attacks targeting other vectors are also possible.}

{One vector for attack is for the service provider to attempt to place cookies or third-party tracking content on the web pages viewed by a user.  Within the EU, the GDPR rules require that users be explicitly informed of such actions and must take a positive step to opt in. Hence attempts at such tracking seem like a relatively minor concern. Outside the EU, existing tools for blocking third-party trackers can be used, leaving the setting of unique identifying first party cookies as the main concern. This can be mitigated by standard approaches e.g. by activists maintaining lists of cookies that can be safely used (similar to existing lists of malware sites, trackers and so on) and users blocking the rest.  }

{Another possible vector of attack is to record the IP address of the user browser, and thereby try to link the ratings back to the individual user. However, due to the widespread use of techniques such as VPN or NAT, use of IP addresses as identifiers is unreliable. Users also have the option of using tools such as TOR to further conceal the link between the IP address revealed to the server and the users identity. Such tools are the subject of an extensive literature in their own right and are complementary to the present discussion.}

The parties here are sometimes referred to as \emph{observers}, rather than attackers, since the relationships here are not fundamentally adversarial being rather of the honest but curious variety. Since our main interest is in honest but curious attackers we exclude active attacks against the UI and user devices from consideration, which are, of course, the subject of an extensive literature in its own right. 
\section{The 3PS Architecture}
\label{sec:3ps:arch} 
\begin{figure}[!ht]
  \caption{Users with proxy agent pool setup\label{fig:system:overview:1}}
  \vspace{2mm}
  \centering 
  \includegraphics[width=0.9\columnwidth]{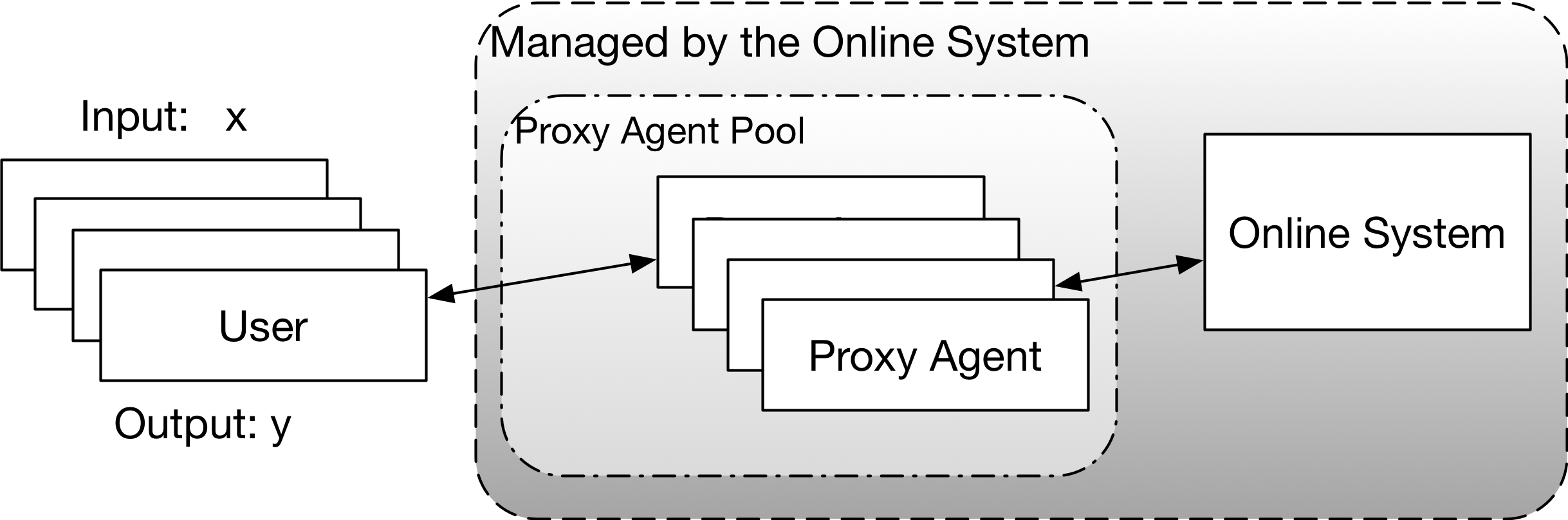}
\end{figure}

The challenge is to construct an online system which satisfies Definition~\ref{def:plausible:deniability}, thereby providing $\delta$-plausible deniability to users, while also providing an effective personalised service. 
We propose an architecture, which we refer to as \emph{3PS}, whereby users access the system through a pool of group identities referred to as \emph{proxy agents}.   This is illustrated schematically in Figure~\ref{fig:system:overview:1}.  The 3PS architecture therefore consists of three interacting parties denoted $\{\UU{}, \PP{}, \S{}\}$ as follows:
\begin{itemize}
	\item An online \emph{system} \S{}. \S{} is a black-box in the sense that only inputs to, and outputs from, \S{} are observable while details of the internal workings of \S{} are hidden from users.
	\item A pool \PP{} of \emph{Proxy Agents} acting as \emph{Group Identities}, routing queries to, and output responses from \S{}. In effect each group identity is an account used to access the system, with this account being shared by multiple users.
	\item A pool \UU{} of  \emph{users} who can submit input to, and receive corresponding output responses from, \S{} via the group identities provided by the proxy agents in \PP{}.  
\end{itemize}%
%
In the 3PS architecture the proxy agent pool is controlled by the backend service. One key reason for doing this is to ensure that proxy agent IDs are recognised as genuine users by the backend system. If not recognised as bona fide users the proxy agents may be flagged as a bot or robot and so trigger defences, such as ``captchas'', or even be blocked. Other than acknowledging the proxy agents as legitimate users, the 3PS system is intended to be backwards compatible and does not require significant engineering changes in  the backend system.
\subsection{Providing Personalisation}
\begin{figure}[!ht]
\caption{Examples of Google Search adverts for individual and shared user profiles.\label{fig:adverts:page}}
  \centering 
  \subfloat[Google Search Adverts for an individual user\label{fig:adverts:page:1}]{%
  \includegraphics[width=0.9\linewidth]{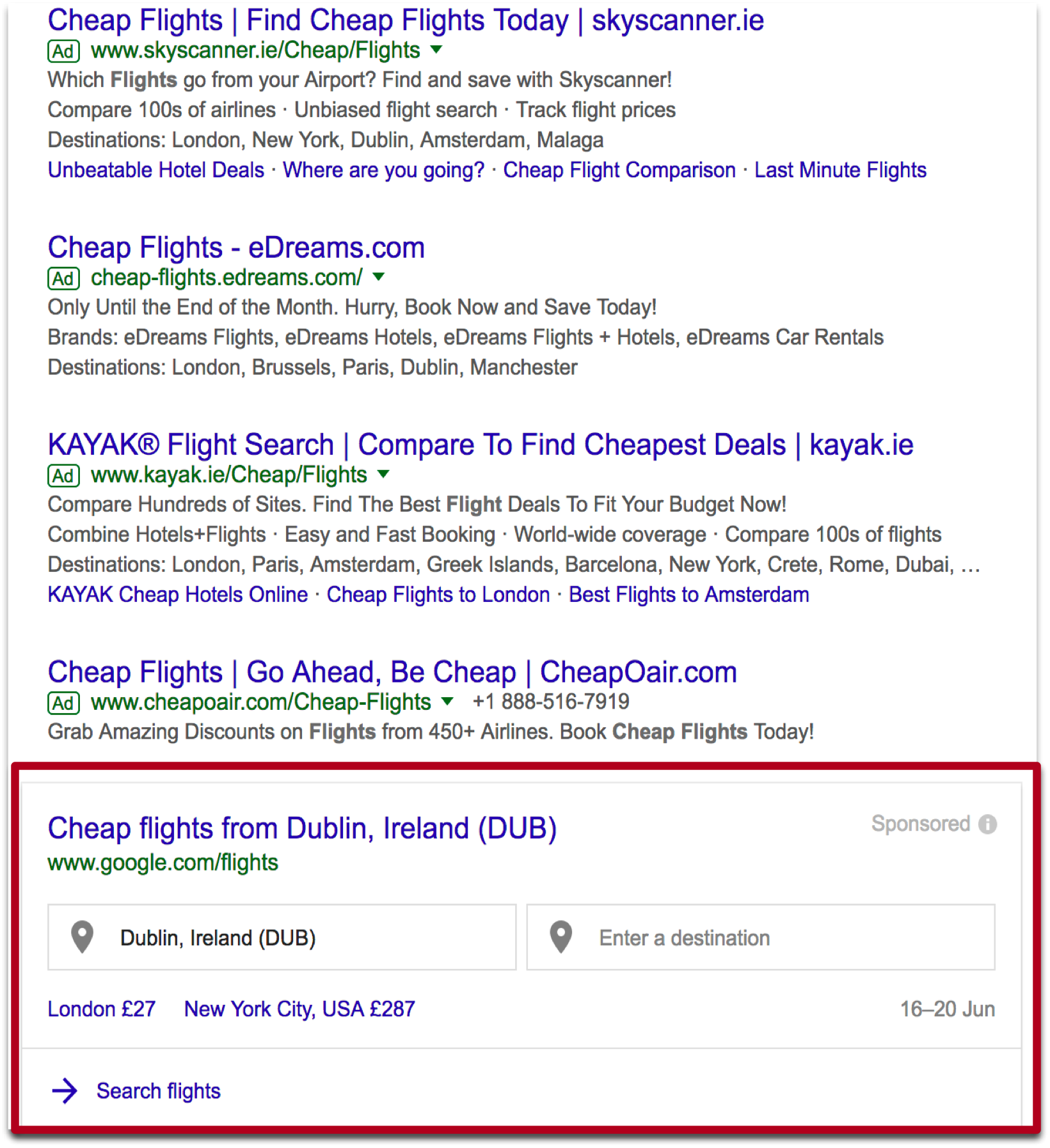}
  }\\[2mm]
 
  \subfloat[Google Search Adverts for a shared proxy user\label{fig:adverts:page:2}]{%
  \includegraphics[width=0.9\linewidth]{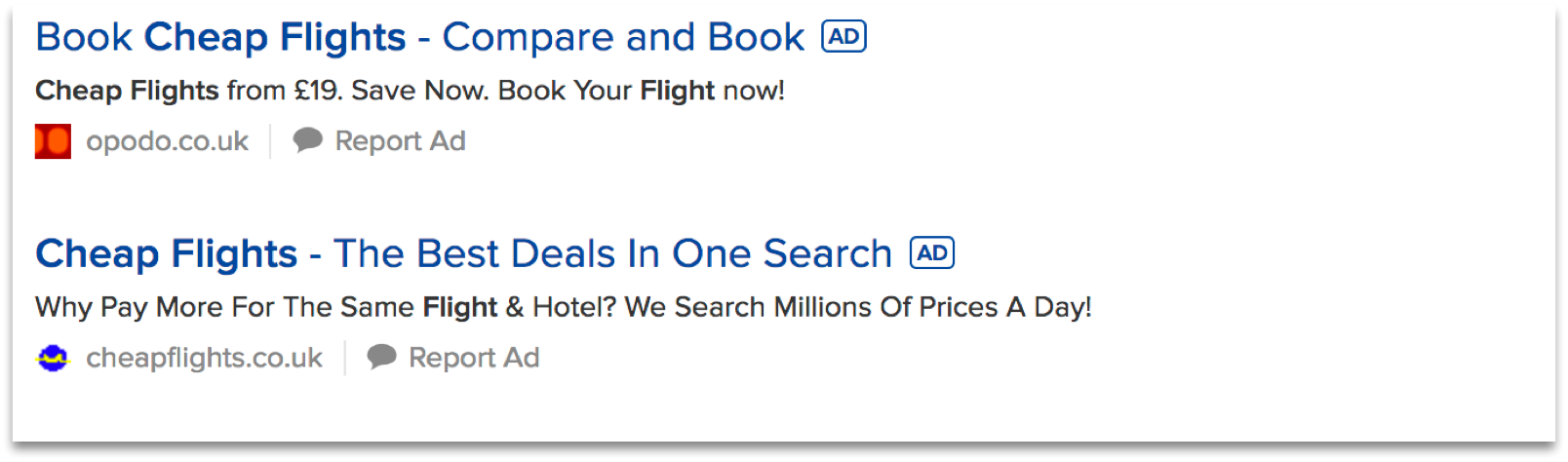}
  } \\[2mm]
\end{figure}

The backend system \S{} is assumed to generate recommendations for a proxy agent based on profiling interests in topics as it would for any other user. In a shared proxy setup users inherit the shared profile of the proxy agent they choose. A user accessing \S{} via the pool of proxy agents and wishing to obtain good recommendations should therefore choose the proxy agent whose interests most closely match their interests.   
 As an example, Figure~\ref{fig:adverts:page:1} and Figure~\ref{fig:adverts:page:2} show the results of issuing the query ``cheap flights'' through two different proxy agent setups. The choice of query is deliberately intended to trigger commercial advertising for illustrative purposes. In Figure~\ref{fig:adverts:page:1} the proxy agent is dedicated to Google Search users located in a single country, Ireland. In Figure~\ref{fig:adverts:page:2} the proxy agent is a web-proxy gateway shared by Google Search users from many countries. 
The response via the proxy agent in Figure~\ref{fig:adverts:page:1} contains significantly more content than the proxy agent in Figure~\ref{fig:adverts:page:2}. Content in Figure~\ref{fig:adverts:page:1} is also more localised to the region of the user, as illustrated by the Google flight search box outlined in red on the figure and in the Ireland ``.ie'' domains on other results. Content obtained from the shared proxy agent in Figure~\ref{fig:adverts:page:2}  by contrast reflects the regional settings of the proxy agent rather than the user -- in this case, UK currency and websites appear in the adverts.

To obtain personalised content, each user chooses a proxy agent closest to their interests in the sense that it is a solution to%

{\footnotesize
\begin{align}
&\min_{p\in\PP{}}\sum_{c\in\SET{C}}| \P(z\in\SEQ{Z}^{u,c}_{u,k}|z\in\SEQ{Z}_{u,k}) - \P(z\in\SEQ{Z}^{u,c}_{u,k} \vert z\in\SEQ{Z}_{p,k}) |
\notag \\
&\qquad\text{s.t. } \quad  \P(z\in\SEQ{Z}^{u, c}_{k} \vert z\in\SEQ{Z}_{p,k}) \leq \delta
\label{eqn:personalise:3}
\end{align}
\normalsize}%
where $\SEQ{Z}_{p,k}$ denotes the input--output interactions of all users with proxy $p$. 
The constraint in \eqref{eqn:personalise:3} ensures that $\delta$-plausible deniability holds for an observer with access to $\SEQ{Z}_{p,k}$.
\subsection{Threat Models}
By varying the observations, $\SEQ{Z}_{att, k}$, available to an observer it is possible to model classes of attack encompassing the system itself and observers with access to more localised background knowledge. We introduce two observer classes we will use in the remainder of this paper.
\subsubsection{Privacy Against A Global Observer} 
\label{Sec:global:observer}
A \emph{global observer} denotes an attacker where $\SEQ{Z}_{att, k}=\SEQ{Z}_{k}$.  That is, with access to all of the input--output interactions for the entire system up to the present step $k$.  A global observer does not have knowledge of the user labelling function $l_u$ but can try to cluster the observed input--output interactions to infer topics of likely interest. This class of attacker encompasses the system itself, external parties such as advertising partners and attackers obtaining data by hacking of the system. 
Provided \eqref{eqn:define:plausible:deniability:0} holds for $\SEQ{Z}_{att,k} = \SEQ{Z}_{k}$ then a user has $\delta$-plausible deniability against global observers.
\subsubsection{Privacy Against A Proxy Observer}
We also consider a \emph{proxy observer}, namely a global observer who also has knowledge of the set of proxy agents $\PP{}_u\subset \PP{}$ used by user $u$.   Hence, a proxy observer knows that the input--output interactions  $\SEQ{Z}_{u,k}$ generated by user $u$ are contained in the subsequence 
\begin{align}
\SEQ{Z}_{att, k}=(z\in\SEQ{Z}_{k} \,: \, \iota_p(z)=1, p\in\PP{}_u)
\label{eq:proxyatt}
\end{align}
where indicator function $\iota_p$ equals $1$ for input--output interactions submitted via proxy $p$ and $0$ otherwise.  From Observation~\ref{prop:1}, a proxy observer is a more powerful attacker than a global observer by having access to more localised data. Provided \eqref{eqn:define:plausible:deniability:0} holds with $\SEQ{Z}_{att, k}$ given by \eqref{eq:proxyatt} then a user has $\delta$-plausible deniability against proxy observers.
\subsection{Mitigating Sybil Attacks}
Attacks by dishonest users who submit false inputs in an attempt to manipulate the outputs of the system are outside the scope of the present paper.  Although this is an important challenge for all online systems it is not specific to 3PS.  That said, the use of shared proxies and unlink-ability of input--output interactions to individual users does potentially facilitate Sybil attacks and so we briefly describe one mechanism, based on the work of~\cite{chaum1988untraceable}, where such attacks can be disrupted while being compatible with the 3PS setup.  In summary, each user mints a number of session tokens (with associated serial number), blinds them with a secret blinding factor and forwards them to the 3PS system through a non-secure channel.  The number of tokens available to a user is limited e.g. by requiring users to authenticate or make payment to the service in order to forward a token, or perhaps by limiting the number of tokens allowed within a certain time window.  Note that during this phase the user might be identified to the system, e.g. to make a payment.  The system then signs the tokens with its private key, without knowledge of the serial number associated with the tokens.  On receiving the signed tokens back from the system, the user can remove the blinding factor and use the tokens to submit inputs to the system anonymously.   Double use of tokens is prevented by the system maintaining a database of the serial numbers of all tokens that have been issued.
\section{Prototype Implementation}
\label{sec:implement}
In this section we describe an experimental implementation of a backend recommender system accepting text queries as inputs and producing text-based outputs.  It is not intended to be a fully working system but rather a proof of concept implemented as software that is sufficient to demonstrate the feasibility of 3PS and to illustrate how personalisation and privacy verification might be implemented.  
In the prototype implementation the internal state of simulated users, proxy agents and the backend system can be inspected for measurement during test. This allows us to conveniently compare probability estimators during experiments that would be private in a production system.
\subsection{Personalisation}
\label{sec:personalisation}
In the prototype implementation inputs and outputs are sequences of words and the dictionaries, $\DICT{X}:=\{\theta^X_1,\theta^X_2,\dots\}$ and $\DICT{Y}:=\{\theta^Y_1,\theta^Y_2,\dots\}$, consisting of common keywords appearing in the input and output respectively. We adopt a standard bag--of--words language model~\cite{manning1999foundations} where features in an input--output pair are modelled as being drawn independently, with replacement, and ignoring order, according to the mixture model 

{\begin{footnotesize}
\begin{align}
& \P(z\in\SEQ{A}_{k} \vert z\in\SEQ{B}_{k}) 
\notag \\
& =\sum_{i=1}^{\vert\DICT{X}\vert} \sum_{j=1}^{\vert \DICT{Y} \vert}
\P( z\in \SEQ{A}_{k} \vert \{\theta^{X}_{i}, \theta^{Y}_{j}\} \in z) 
\P( \{\theta^{X}_{i}, \theta^{Y}_{j}\} \in z\vert  z\in \SEQ{B}_{k})
\label{eqn:bag:of:words:discrete:1}
\end{align}
\end{footnotesize}}%
where $\SEQ{A}_{k}, \SEQ{B}_{k} \subseteq \SEQ{Z}_{k}$ are non-empty sequences of observations, \cite{Hofmann:1998:SMC:888741}. The quantity $\P( z\in \SEQ{A}_{k} \vert \{\theta^{X}_{i}, \theta^{Y}_{j}\} \in z) $ is the probability that an input-output pair $z$ belongs to subsequence $\SEQ{A}_{k}$ given the keywords $\{\theta^{X}_i, \theta^{Y}_j\}$ co-occur in $z$. Similarly $\P( \{\theta^{X}_{i}, \theta^{Y}_{j}\} \in z\vert  z\in \SEQ{B}_{k})$ is the probability that keywords $\{\theta^{X}_i, \theta^{Y}_j\}$ co-occur in $z$ given that $z$ belongs to subsequence $\SEQ{B}_{k}$.

Expression \eqref{eqn:bag:of:words:discrete:1} can be applied directly to \eqref{eqn:personalise:3} so that

{\footnotesize
\begin{align}
&\P(z\in\SEQ{Z}^{u, c}_{u,k}|z\in\SEQ{Z}_{u,k}) - \P(z\in\SEQ{Z}^{u, c}_{u,k} \vert z\in\SEQ{Z}_{p,k}) \notag \\
&= \sum_{i=1}^{\vert\DICT{X}\vert} \sum_{j=1}^{\vert \DICT{Y} \vert}\underbrace{\P(z\in\SEQ{Z}^{u, c}_{u,k}\vert \{\theta^{X}_{i}, \theta^{Y}_{j}\} \in z)}_{(a)} \notag \\
& \qquad \times	\left( \underbrace{\P(\{\theta^{X}_{i}, \theta^{Y}_{j}\} \in z \vert z\in  \SEQ{Z}_{u,k})}_{(b)} - \underbrace{\P(\{\theta^{X}_{i}, \theta^{Y}_{j}\} \in z \vert z\in  \SEQ{Z}_{p,k})}_{(c)} \right)
\label{eqn:personalise:4}
\end{align}
\normalsize}%
and the minimisation element of \eqref{eqn:personalise:3} becomes a calculation over the term labelled (c) in \eqref{eqn:personalise:4}. We will return to the constraint element of \eqref{eqn:personalise:3} later.

Term \eqref{eqn:personalise:4}(a) is the only element of the RHS of  \eqref{eqn:personalise:4} that depends on knowledge of the user labelling function $l_u$. Since \eqref{eqn:personalise:4}(a) and \eqref{eqn:personalise:4}(b) do not depend on $\SEQ{Z}_{p,k}$ they can be estimated privately by $u$. To allow \eqref{eqn:personalise:4} to be \emph{privately} by a user,  it is sufficient for each proxy agent $p\in\PP{}$ to release the probability distribution \eqref{eqn:personalise:4}(c) \emph{publicly}. With this a user can construct \eqref{eqn:personalise:4}. 

Expression \eqref{eqn:personalise:4} consists of matrix multiplications of matrices of size $\vert\DICT{X}\vert\times\vert\DICT{Y}\vert$. The proxy selection condition in \eqref{eqn:personalise:3} can be solved efficiently in practice by estimating the various probabilities. 
\subsection{Estimating Probabilities}
\label{sec:estimating:probabilities}
To estimate probabilities in our prototype implementation, user $u$ applies their private labelling function $l_u$ to label each input--output pair $\{x, y\}\in\SEQ{Z}_{u,k}$ for topics in $\SET{C}$. 
Let $\SEQ{U}^{c}_{u,k}$ and $\SEQ{V}^{c}_{u,k}$ denote the labelled inputs and outputs of $\SEQ{Z}^{u,c}_{u,k}$ respectively.  Apply count-vectorisation to each element of $\SEQ{U}^{c}_{u,k}$ and $\SEQ{V}^{c}_{u,k}$ and gather the result into count-matrices $\MAT{A}_{c}$ and $\MAT{B}_{c}$ of size $\vert \SEQ{U}^{c}_{u,k} \vert \times \vert \DICT{X}\vert$ and $\vert \SEQ{V}^{c}_{u,k} \vert \times \vert \DICT{Y}\vert$ respectively. 
Since $\vert \SEQ{U}^{c}_{u,k} \vert = \vert \SEQ{V}^{c}_{u,k} \vert$, the quantity $\MAT{N}_{c} = \MAT{A}_{c}^{T}\MAT{B}_{c}$ is of dimension $\vert \DICT{X}\vert \times \vert \DICT{Y}\vert$.  $\MAT{N}_{c}$ is the count co-occurrence matrix of input--output interactions of input--output features in $\SEQ{Z}_{u,k}$ labelled for topic $c$. The $ij$--element  of matrix $\MAT{N}_{c}$, denoted $\MAT{N}_{c, ij}$, is the co-occurence count of the features $\{\theta^{X}_{i}, \theta^{Y}_{j}\}$ in $\SET{Z}_{u,k}$ labelled for topic $c\in\SET{C}$.  
We apply regular Laplace Smoothing, \cite{Manning:2008:IIR:1394399}, to avoid divide by zero underflows in subsequent computations when there are sparse occurrences of keywords in $\SEQ{Z}_{u,k}$. Laplace smoothing resolves this problem by adding a factor $\lambda_u > 0$ to each keyword count  so that $N_{c,ij} \longrightarrow N_{c,ij} + \lambda_u$. 
The quantity

\begin{align}
&\PHAT(\{\theta^{X}_{i}, \theta^{Y}_j\} \in z \vert z\in \SEQ{Z}^{u,c}_{u,k}) = \frac{N_{c,ij}}{ N_{c}} 
\notag\\  
&N_{c}  = \sum_{i=1}^{\vert\DICT{X} \vert} \sum_{j=1}^{\vert\DICT{Y} \vert} N_{c,ij}
\label{eqn:coocurrence:base}
\end{align}
is then an estimator for $\P(\{\theta^{X}_{i}, \theta^{Y}_j\} \in z \vert z\in \SEQ{Z}^{u,c}_{u,k})$.    
Similarly, an estimator for $\P(\{\theta^{X}_{i}, \theta^{Y}_{j}\} \in z \vert z\in  \SEQ{Z}_{u,k})$ is given by 

{\footnotesize
\begin{align}
	&\PHAT(\{\theta^{X}_{i}, \theta^{Y}_{j}\} \in z \vert z\in  \SEQ{Z}_{u,k}) 
		= \frac{N_{ij}}{N}
	\label{eqn:coocurrence:c} 
	\\
	&N = \sum_{c\in\SET{C}}N_c, \quad N_{ij} = \sum_{c\in\SET{C}}N_{c,ij} \notag  
\end{align}
}\normalsize
and 
\begin{align}
\PHAT(z\in\SEQ{Z}^{u,c}_{u,k} \vert z\in  \SEQ{Z}_{u,k}) = \frac{N_c}{N}
\label{eqn:prob:c} 
\end{align}%
is an estimator for the probability of an observation being labelled for topic $c$. 

Let $\MAT{O}$ have components $O_{ij}(z)$ given by
\begin{align*}
O_{ij}(z) =
\begin{cases}
1 \text{ if } \phi^{X}_{i}(x) > 0 \text{ and }\phi^Y_{j}(y) > 0 \text{ for } z=\{x,y\} \\
0 \text{ otherwise }	
\end{cases}	
\end{align*}
and define 
\begin{align*}
O_{c,ij} := \sum_{z\in\SEQ{Z}^{u,c}_{u,k}}
&O_{ij}(z), \qquad  
O_{c} := \sum_{i=1}^{\vert\DICT{X} \vert} \sum_{j=1}^{\vert\DICT{Y} \vert} O_{c,ij} \\
&\text{and, } O := \sum_{c\in\SET{C}} O_{c}
\end{align*} 
so that an estimator for $\P(z\in\SEQ{Z}^{u,c}_{u,k}\vert \{\theta^{X}_{i}, \theta^{Y}_{j}\} \in z)$ is 
%
\begin{align}
&\PHAT(z\in\SEQ{Z}^{u,c}_{u,k}\vert \{\theta^{X}_{i}, \theta^{Y}_{j}\} \in z)	
= \frac{O_{c,ij}}{O_{c}}
	\label{eqn:coocurrence:conditional:c}
\end{align}
and an estimator for $\P(z\in\SEQ{Z}_{u,k}\vert \{\theta^{X}_{i}, \theta^{Y}_{j}\} \in z)$ 
\begin{align}
&\PHAT(z\in\SEQ{Z}_{u,k}\vert \{\theta^{X}_{i}, \theta^{Y}_{j}\} \in z)	
= \frac{\sum_{c\in\SET{C}}O_{c,ij}}{O}	\label{eqn:coocurrence:conditional:c2}
\end{align}

For a proxy agent $p$, let $\SEQ{U}_{p,k}$ and $\SEQ{V}_{p,k}$ denote the inputs and outputs in $\SEQ{Z}_{p,k}$ respectively. Apply count-vectorisation to each element of $\SEQ{U}_{p,k}$ and $\SEQ{V}_{p,k}$ and gather the result into count-matrices $\MAT{C}$ and $\MAT{D}$ respectively of size $\vert \SEQ{U}_{p,k} \vert \times \vert \DICT{X}\vert$ and $\vert \SEQ{V}_{p,k} \vert \times \vert \DICT{Y}\vert$ respectively. The quantity $\MAT{M} = \MAT{C}^{T}\MAT{D}$, of dimension $\vert \DICT{X}\vert \times \vert \DICT{Y}\vert$, is the count co-occurrence matrix of input--output interactions of input--output features in $\SEQ{Z}_{p,k}$, to which Laplace smoothing is applied.
We estimate  $\P(\{\theta^{X}_{i}, \theta^{Y}_j\} \in z \vert z\in \SEQ{Z}_{p,k})$ for each proxy agent $p$ as
{
\begin{align}
&\PHAT(\{\theta^{X}_{i}, \theta^{Y}_j\} \in z \vert z\in \SEQ{Z}_{p,k}) = \frac{M_{ij}}{M }, 
\,\, M  = \sum_{i=1}^{\vert\DICT{X} \vert} \sum_{j=1}^{\vert\DICT{Y} \vert} M_{ij}
\label{eqn:coocurrence:0}
\end{align}
}%
and  $M_{ij}$ denotes the $ij$--element  of matrix $\MAT{M}$.  

Expressions \eqref{eqn:coocurrence:c}, \eqref{eqn:coocurrence:conditional:c} and \eqref{eqn:coocurrence:0} can then be combined, to estimate the RHS of \eqref{eqn:personalise:4} for each user $u$. 

In our experimental setup, it is convenient to estimate plausible deniability directly from the definition \eqref{eqn:define:plausible:deniability:0} as 
\begin{align}
\PDD{} :=
\PHAT(z\in\SEQ{Z}_{k}^{u,c} \vert z\in\SEQ{Z}_{att,k}) 
=
\frac{\vert z\in\SEQ{Z}_{att,k} \, : \, l_u(z) = c \vert}
{\vert z\in\SEQ{Z}_{att,k} \vert}
\label{eqn:experiment:estimate:pd:direct}
\end{align}%
The probability of user $u$ observing an input--output pair labelled with topic $c$ when accessing \S{} through proxy agent $p$ is 
$\P(z\in\SEQ{Z}^{u,c}_{p,k} \vert z\in\SEQ{Z}_{p,k})$. This is estimated in our experimental setup as
\begin{align}
\PHAT(z\in\SEQ{Z}^{u,c}_{p,k} \vert z\in\SEQ{Z}_{p,k}) 
=
\frac{\vert z\in\SEQ{Z}_{p,k} \, : \, l_u(z) = c \vert}
{\vert z\in\SEQ{Z}_{p,k} \vert}	
\label{eqn:eval:utility:2}
\end{align}%
and $\P(z\in\SEQ{Z}^{u,c}_{u,k} \vert z\in\SEQ{Z}_{u,k})$, the probability of user $u$ observing an input--output pair labelled with topic $c$ when accessing \S{} directly is estimated as
\begin{align}
\PHAT(z\in\SEQ{Z}^{u,c}_{u,k} \vert z\in\SEQ{Z}_{u,k}) 
=
\frac{\vert z\in\SEQ{Z}_{u,k} \, : \, l_u(z) = c \vert}
{\vert z\in\SEQ{Z}_{u,k} \vert}	
\label{eqn:eval:utility:2a}
\end{align}

We measure the estimated \emph{utility loss} incurred by user $u$ as a result of selecting proxy agent $p$, using \eqref{eqn:eval:utility:2} and \eqref{eqn:eval:utility:2a}, as

{\footnotesize
\begin{align}
&\Delta U_{p,k} ^{u,c} :=
\frac{1}{2}\sum_{c\in\SET{C}}\vert \PHAT(z\in\SEQ{Z}^{u,c}_{u,k} \vert z\in\SEQ{Z}_{u,k}) 
-
\PHAT(z\in\SEQ{Z}^{u,c}_{p,k} \vert z\in\SEQ{Z}_{p,k}) \vert 
\label{eqn:eval:utility:1}
\end{align}
\normalfont}%
that is, the total variation between the sensitive topic probability estimator the user would calculate if they used \S{} directly and the probability estimator of the topic calculated by the proxy agent they used.
\subsection{User Estimate of Privacy Threat}
\label{sec:testing:plausible:deniability}
The challenge for a user in checking \eqref{eqn:define:plausible:deniability:0}  is that it requires knowledge of  $\SEQ{Z}^{u,c}_{k}$ by user $u$. So that $u$ is required to know the history of input--output interactions for each sensitive topic $c$ for \emph{all} users in the 3PS system. 

In the prototype implementation we use the approach that each user $u$ has defined a set, $\THA{} \subseteq \DICT{X} \times \DICT{Y}$, for each sensitive topic $c$, consisting of input--output keywords whose presence means an input--output observation is labelled as sensitive by $u$. In experiments, \THA{}, is selected for each user $u$ and topic $c$ using the training data to choose the keyword pairs for which 
\begin{align}
\THA(\alpha) =
\left\{ \{\theta^{X}_i, \theta^{Y}_j\} \, : \,  \{\PHAT(z\in\SEQ{Z}_{u,k}^{u,c} \vert \{\theta^{X}_i, \theta^{Y}_j\} \in z) > \alpha \right\}
\label{eqn:eval:pd:1}
\end{align}%
where $0 <\alpha\leq 1$ is a parameter chosen using cross-validation. 

For each topic $c$ define the associated indicator function over observations $z\in \SEQ{Z}_{k}$ and $\{\theta^{X}_i, \theta^{Y}_j\} \in \THA{}(\alpha)$, as
\begin{align}
\iota_{\alpha}^{c} (\{\theta^{X}_i, \theta^{Y}_j\} \vert z) = 
\begin{cases}
	1 \text{ if } 
	\{\theta^{X}_i, \theta^{Y}_j\} \in z
	\\
	0 \text{ otherwise}
\end{cases}	
\label{eqn:verify:pd:2}
\end{align}%
That is, the indicator function labels an observation as sensitive if it contains an input--output keyword pair from $\THA(\alpha)$ and non-sensitive otherwise. 
Using the bag-of-words model to combine this with the published estimator $\PHAT( \{\theta^{X}_{i}, \theta^{Y}_{j}\} \in z\vert  z\in \SEQ{Z}_{p,k})$ provided by each proxy agent we get an estimator for  $\P(z\in\SEQ{Z}^{u,c}_{k} \vert z\in\SEQ{Z}_{p,k})$ given by

{\footnotesize
\begin{align}
&\PHAT_{\alpha}(z\in\SEQ{Z}^{u,c}_{k} \vert z\in\SEQ{Z}_{p,k}) = \notag\\
&
\frac{
\sum_{i=1}^{\vert\DICT{X}\vert} \sum_{j=1}^{\vert \DICT{Y} \vert}
\iota_{\alpha}^{c}(\{\theta^{X}_i, \theta^{Y}_j\} \vert z)
\PHAT( \{\theta^{X}_{i}, \theta^{Y}_{j}\} \in z\vert  z\in \SEQ{Z}_{p,k})
}{
\sum_{c\in\SET{C}}\sum_{i=1}^{\vert\DICT{X}\vert} \sum_{j=1}^{\vert \DICT{Y} \vert}
\iota_{\alpha}^{c}(\{\theta^{X}_i, \theta^{Y}_j\} \vert z)
\PHAT( \{\theta^{X}_{i}, \theta^{Y}_{j}\} \in z\vert  z\in \SEQ{Z}_{p,k})
}
\label{eqn:verify:pd:3a}
\end{align}
\normalsize}%

In a real-world setup it is up to the user to  decide how to select \THA{}. For example, the \textbf{PRI} tool developed in \cite{mac.P1} and \cite{mac.P2} allows a user to analyse input--output observations for privacy threats and so assess which keyword pairs are more or less revealing of sensitive topics.  In this way tools such as \textbf{PRI} can provide information to assist in constructing \THA{} in a real-world setup.
%
%
\section{Experimental Setup}
\label{sec:privacy:setup}
\subsection{General Setup}
In our experimental setup, the test datasets, described later, are labelled with a set of topics \SET{C}. Before an experimental run each user and proxy agent simulated during the experiment is allocated a topic of interest from \SET{C}.  When a user or proxy agent is allocated the non-sensitive, catch-all topic $c_0$ we will say the user or proxy agent is \emph{randomly initialised} meaning that they have no interest in a specific sensitive topic. We call the percentage of proxy agents in \PP{} or users in \UU{} that have been randomly initialised the \emph{diversity} of \PP{} or \UU{}. During experiments we will typically report   results for $0\%$, $50\%$ and $100\%$ diversity in \PP{} and/or \UU{}. 

At the start of each experimental run, each user and each proxy agent is allocated   initial data consisting of input--output pairs from the test dataset labelled for their allocated topic of likely interest, referred to as \emph{background knowledge}. Each user and proxy agent in the simulation has a copy of the common dictionaries \DICT{X} and \DICT{Y} from \S{}. Next, each user and each proxy agent estimates initial values of the probabilities in Section~\ref{sec:estimating:probabilities} from the initial background knowledge using \DICT{X} and \DICT{Y}. We refer to these probabilities as the \emph{internal state} of the user or proxy agent.  
An input query is a keyword in \DICT{X} drawn from $\THA(\alpha = 0.5)$ at random by $u$.  

Users select a proxy agent best matching their allocated topic of interest by solving \eqref{eqn:personalise:3}.
When a proxy agent receives an input query from a user it passes it directly to \S{}. Since the set of topics is known to \S{} in our experiments, \S{} creates a personalised response by solving $c^{*} = \argmax_{c\in\SET{C}} \PHAT(z\in\SEQ{Z}_{p,k}^{\S{}, c} \vert \{\theta^{X}_i\} \in z)$, to find the topic of maximum likely interest from \SET{C} given the input it received, and then selecting an output labelled for $c^{*}$. The resulting output is returned to the proxy agent. The input--output interaction pair is added to the background knowledge of the proxy agent and its internal state is updated with new probability estimates. The output is routed to the requesting user and the same input--output interaction is added to its background knowledge and its internal state and probability estimator are updated.

Background knowledge is not shared among users and proxy agents.  When a user switches to a different proxy agent during an experimental run, the user history of input--output interactions does not transfer to the new proxy agent so that individual proxy agents see only the history of interactions from users accessing \S{} through it. A full reset is performed between test runs by re-initialising the entire setup.  
\subsection{Data Sources}
Data from three real-world sources are used in experiments.
\begin{description}[style=unboxed,leftmargin=0cm, labelindent=0cm]\itemsep2pt
\item[Hotels] Tripadvisor hotel reviews containing hotel review titles, review bodies and lowest price per room downloaded from, \cite{wang.data}, and consisting of over $1.6$ million hotel reviews. Queries consisting of words extracted from review titles are used as inputs and detailed review bodies represent outputs.
\item[Products] Product review titles, review bodies and overall rating scores downloaded from, \cite{wang.data}, containing  Amazon product reviews for $6$ types of merchandise and consisting of over $2.2$ million product reviews. Words appearing in product review titles are used as query inputs and outputs review bodies.
\item[Search] Web search queries and corresponding result pages relevant to $5$ sensitive topics $\{$``weight loss'', ``anorexia'', ``diabetes'', ``bad credit history'', ``pregnancy''$\}$ used in \cite{mac.P1} `and \cite{mac.P2} and comprising $86,837$ Google searches constructed by gathering search terms from the Wikipedia article related to each sensitive topic and from the top web search queries appearing on \url{www.Soovle.com} for the non-sensitive queries. Here the queries submitted to Google are the inputs with the corresponding result pages taken as outputs.
\end{description} 
\subsection{Assigning Topics}
\label{sec:labelling:topics}
Default topics for experiments were defined as follows from each of the test datasets.
\begin{description}[style=unboxed,leftmargin=0cm, labelindent=0cm]\itemsep2pt

\item[Hotels] Five topic categories are defined by dividing the \emph{lowest price per room} into equally spaced ranged, namely $0:= [0, 110], 1:=(110, 220], 2:=(220, 330], 3:=(330, 440], 4:=(440, 550], 5:=(550,\infty)$.  Reviews are then labeled according to the lowest price.
\item[Products] The \emph{overall rating score} is used to define topic categories, namely very dissatisfied (Topic 1) to very satisfied (Topic 4).  Topic $0$ is used to indicate no rating was given so there are $5$ topic categories in total.
\item[Search] There are $6$ topic categories labelled $\{0:=$ ``Other'', $1:=$``{weight loss}'', $2:=$``{anorexia}'', $3:=$``{diabetes}'', $4:=$``{bad credit history}'', $5:=$``{pregnancy}''$\}$ as in \cite{mac.P1, mac.P2}   Each input--output pair is labelled with the topic the input query refers to. 
\end{description}%
When experiments are performed requiring a larger number of topics than those above, the Hotels dataset is divided into the required number of topic categories by specifying different lowest price ranges. In this way it is possible to create a variety of topic categories automatically by re-grouping the data into finer price categories to create more topic categories. The Hotel dataset was chosen for convenience since the categories are defined by numeric, price-per-room, ranges and so it is straightforward to programatically  define more categories by changing the numeric ranges. 

\subsection{Revealing Keyword Pairs}
Each of the test datasets was preprocessed using the text processing described in Section~\ref{sec:estimating:probabilities} to produce dictionaries $\DICT{X}$ and $\DICT{Y}$ for each dataset. 
A range of dictionary sizes from $50$ to $1000$ features was assessed by selecting random subsequences $\SEQ{A}_{k} \subseteq \SEQ{Z}_{k}$ and choosing the dictionaries that minimise 

{\footnotesize
\begin{align}
\vert &\PHAT(z\in\SEQ{A}_{k}	\vert z\in\SEQ{Z}_{k}) \notag\\
&- \sum_{i=1}^{\vert\DICT{X}\vert} \sum_{j=1}^{\vert \DICT{Y} \vert} \PHAT(z\in\SEQ{A}^{u,c}_{k}\vert \{\theta^{X}_{i}, \theta^{Y}_{j}\} \in z) 
\PHAT(\{\theta^{X}_{i}, \theta^{Y}_{j}\} \in z \vert z\in  \SEQ{Z}_{k})\vert
\end{align}
\normalfont}

From this we selected $\vert\DICT{X}\vert = 250$ and $\vert\DICT{Y}\vert = 500$ for our experiments.
 
\begin{figure}[ht]
\vspace{2mm}
  \centering 
  \includegraphics[width=0.9\columnwidth, height=6cm]{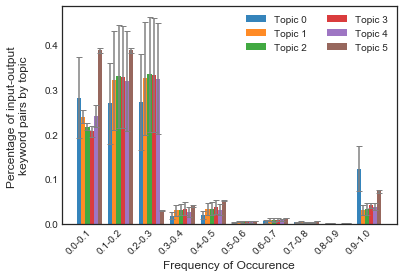}
 \caption{Frequency of co-occurence of keyword pairs by topic averaged over samples from all datasets, sample variation is shown as error per topic\label{fig:estimate:delta:0}}
\end{figure}

The distribution of keyword pairs in samples drawn from each of the three test datasets is shown in Figure~\ref{fig:estimate:delta:0} by topic. Average values were calculated by taking $10$ samples each of $10,000$ items from each of the test datasets. Error bars in Figure~\ref{fig:estimate:delta:0} indicates variance from sampling.  In the case of all datasets and for all topics, the co-occurrence frequency of the majority of keyword  pairs fall below $0.3$. The rarest keyword pairs by topic, and hence the most revealing, have co-occurrence frequencies greater than $0.5$. These keyword pairs comprise less than $10\%$ of the total keyword pairs, suggesting that the most revealing keyword pairs form a small subset in the case of all datasets. 
%

%
%
%
\section{Experimental Evaluation}
\subsection{Topic Diversity and User Numbers}
\label{sec3ps:eval:diversity}
We assess the effects of topic diversity and user numbers for the case consisting of a single proxy agent and a single sensitive topic. We denote the senstive topic $c_1$ so that $\SET{C} := \{c_0, c_1\}$ where $c_0$ is the catch-all topic. A single proxy agent setup means $\SEQ{Z}_{k} := \SEQ{Z}_{p,k}$ so that results here apply to both proxy and global observers.
\begin{figure}[!ht]
\vspace{2mm}
  \centering 
  \includegraphics[width=\columnwidth]{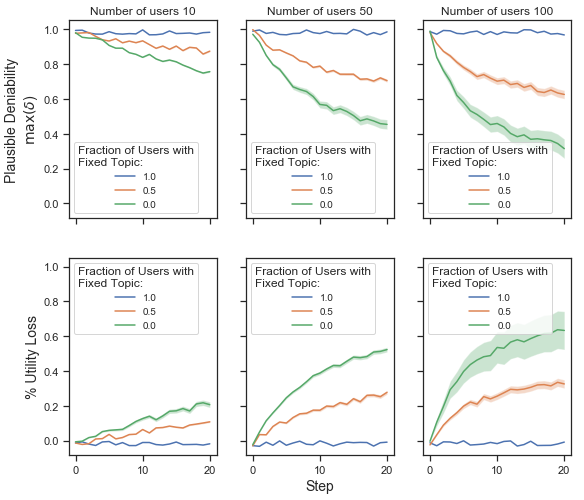}
  \caption{Effect of topic diversity among users on plausible deniability and utility loss for a single proxy agent with initial fixed topic interest by user diversity and number of users (A step is an input--output pair event)\label{fig:system:testing:plausible deniability:0}}
  \end{figure}%
Tests were repeated with $0\%, 50\%$ and $100\%$  of users having  $c_u = c_0$ and the remainder having $c_u = c_1$. We report results for $10$, $50$ and $100$ users for compactness. Results are averaged by dataset and error about the mean is shown as a shaded region.
Plausible deniability, from \eqref{eqn:experiment:estimate:pd:direct}, and utility loss, from \eqref{eqn:eval:utility:1}, averaged over users, are shown in Figure~\ref{fig:system:testing:plausible deniability:0}. Plausible deniability is plotted in the first row and utility loss in the second row.  

From \eqref{eqn:define:plausible:deniability:0}, a user has better plausible deniability for lower values of $\delta$ since $\delta$ is an upper bound.
Our results suggest that increasing user numbers decreases $\delta$ and so \emph{improves} plausible deniability but \emph{only} when users have varied interests. Once users have a diverse range of interests, increasing the number of users is observed to accelerate improvement in plausible deniability.  For utility loss, increasing volumes of users without specific interests is observed to increase utility loss. When all users of a proxy agent have no specific topic interests so that diversity is high this is reflected in increased utility loss relative to topic $c_1$ as one might expect. 

%
\subsection{Personalisation Performance}
\label{sec:proxy:choice}
\begin{figure}[!ht]
\vspace{2mm}
  \centering 
  \includegraphics[width=\columnwidth]{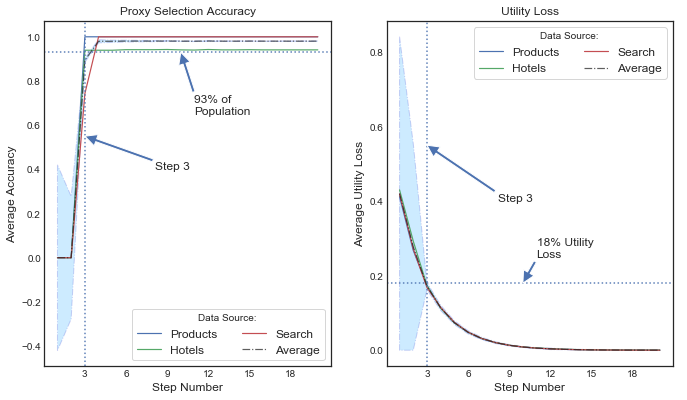}
   \caption{User to Proxy Agent Selection Accuracy (LHS) and Utility Loss (RHS) averaged over all experimental datasets\label{fig:system:cluster:0}}
  \end{figure}

In 3PS users select proxies closest to their interests but the responses generated by proxy agents also change as users submit queries via them.  We would like this joint selection/update process to converge so as to achieve good personalisation performance.  In this section we use our prototype implementation to evaluate this process.  
Experimental setups with proxy pools of sizes $3\leq\vert\PP{}\vert\leq30$ and numbers of users $10\leq\vert\UU{}\vert \leq 120$ were configured for each of the test datasets. 
We
initialise proxy agents in $\PP{}$ randomly so that there is no automatic choice of best proxy agent--user match. 
Users are allocated a sensitive topic as their target topic from the set of topics in each of the test datasets. Each user applies \eqref{eqn:personalise:3} to select a proxy agent best matching their target topic by enumerating each proxy agent in $\PP{}$ in turn. Users only submit queries related to the their allocated topic of interest so that noise due to diverse topic interests of users is controlled in the setup here to focus on convergence properties. 
Once a proxy agent is selected a user issues a query related to their topic of interest and the internal states of users and proxy agents are updated accordingly.
Results are reported as averages over $\vert\PP{}\vert$ and $\vert\UU{}\vert$ and topic for compactness and shown in Figure~\ref{fig:system:cluster:0}.

The measured accuracy of \eqref{eqn:personalise:3} for proxy agent selection is shown in the LHS plot of Figure~\ref{fig:system:cluster:0}. 
Proxy agent selection is deemed to be accurate when a user chooses a proxy agent whose allocated topic of most likely interest matches the allocated target topic of the user. 
The RHS of Figure~\ref{fig:system:cluster:0} is the utility loss, calculated from \eqref{eqn:eval:utility:1}, taken at each input--output step. For visual clarity, standard error is shown for the average utility loss over all datasets.
Utility loss is high and accuracy is low initially reflecting the fact that the initial internal state of proxy agents is randomly set.
Convergence to the proxy agent with closest interests is observed to happen quickly for all data sources, achieving at least $93\%$ accuracy for all datasets after $3$ iterations with a corresponding average utility loss of $20\%$. When averaged over all data sources the average accuracy is $98\%$ after $3$ input--output steps. The utility loss is also observed to decrease for all topics over time, reaching an average across all datasets of $0.18$ after $3$ input--output iterations and $0.0002$ by iteration $20$.  
 
Users are observed to select the correct proxy agent with greater than $90\%$ accuracy, and to reject all proxy agents with $100\%$ accuracy if there is no suitable proxy agent available. Overall, in experiments where the ratio of users to proxy agents was increased from $1:1$ to $30:1$, the utility loss is observed to decrease more slowly as the average number of users attaching to each proxy agent increases. When the ratio of users to proxy agents was $30:1$, for example, the average utility loss on step 1 was $0.67$. Convergence to a low utility loss was also observed to be rapid, even at high user to proxy agent ratios, reaching $0.18 \pm 0.02$ after 4 input--output steps when the user to proxy agent load factor was $30:1$.

The number of topic categories was also varied by regrouping the Hotel dataset.  High proxy agent selection accuracy was consistently observed, with accuracy of greater than $90\%$ after step 3. The utility loss was also  observed to decrease rapidly to less than $0.20 \pm 0.02$ after 4 input--output steps, reaching  minimum of less than $0.01$ by iteration $20$ on average over all topics.

Overall, the results suggest that the proxy agent selection method converges rapidly and accurately, providing a high degree of personalisation. Utility loss also decreases rapidly as more topic specific input--output events are observed. This is consistent across the test datasets, and for a range of user--to--proxy agent ratios, suggesting that the proxy agent selection mechanism performs well across a variety of setups.
\subsection{Plausible Deniability}
\label{sec:test:pd}
\begin{figure}[!ht]
\vspace{2mm}
  \centering 
  \includegraphics[width=\columnwidth, height=6cm]{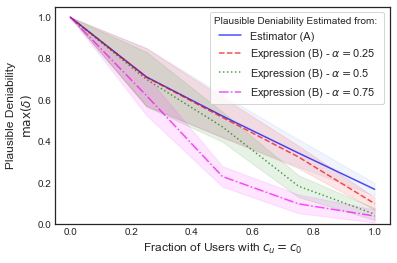}
  \caption{Plausible deniability by topic averaged over all datasets, topics, sizes of proxy agent pool and number of users. Expression (A) indicates use of  \eqref{eqn:experiment:estimate:pd:direct}, and Expression (B) use of  \eqref{eqn:verify:pd:3a} with value of $\alpha$ shown.\label{fig:estimate:delta:10}}
\end{figure}

\begin{figure}[!ht]
\vspace{2mm}
  \centering 
  \includegraphics[width=\columnwidth, height=6cm]{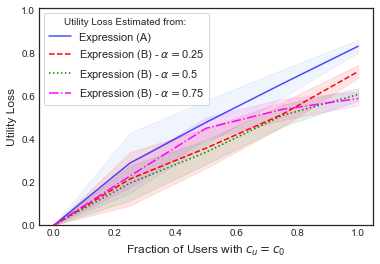}
  \caption{Utility Loss averaged over all datasets, topics, sizes of proxy agent pool and number of users. Expression (A) indicates use of  \eqref{eqn:experiment:estimate:pd:direct}, and Expression (B) use of  \eqref{eqn:verify:pd:3a} with value of $\alpha$ shown.\label{fig:estimate:delta:11}}
\end{figure}

We next assess the degree of plausible deniability protection available to users with respect to a proxy observer when there are multiple proxy agents. We also assess how diversity in user topic interests influences plausible deniability and utility loss. 
Since a proxy observer is at least as powerful as a global observer the results here provide worst-case bounds in the face of a global observer. 
Experimental setups with proxy pools of sizes $3\leq\vert\PP{}\vert\leq30$ and numbers of users $10\leq\vert\UU{}\vert \leq 120$ were configured for each of the test datasets. 
Each proxy agent $p\in\PP{}$ was allocated a topic $c_{p} \in \SET{C}$ as their topic of interest. 
Each user $u\in\UU{}$ was allocated with a target topic of interest $c_u\in\SET{C}$ with setups of $0\%, 25\%, 50\%, 75\%$ and $100\%$ of users having $c_u = c_0$ to model various levels of diversity of topic interests among users. Results are reported as averages over $\vert\PP{}\vert$ and $\vert\UU{}\vert$ and topic for compactness and shown in Figure~\ref{fig:estimate:delta:10} and Figure~\ref{fig:estimate:delta:11}.

In Figure~\ref{fig:estimate:delta:10} we show measurements of estimated level of plausible deniability.  
 We show estimates of \PDDPROXY{} calculated directly from  \eqref{eqn:experiment:estimate:pd:direct}, together with the values of the estimator \eqref{eqn:verify:pd:3a} calculated using $\THABASE{}(\alpha)$  as the set of sensitive keywords. To model the situation where the user has  partial or censored dictionaries \DICT{X} and \DICT{Y} in experiments, we show measurements for values for $\alpha \in \{0.25, 0.5, 0.75\}$. 

The results shown in Figure~\ref{fig:estimate:delta:10} indicate that plausible deniability is observed to improve monotonically as diversity of user interest in topics increases. This is true when either  \eqref{eqn:experiment:estimate:pd:direct} or \eqref{eqn:verify:pd:3a} are used as estimators, for all values of $\alpha$.
The estimated value using \eqref{eqn:verify:pd:3a} is consistently lower than the corresponding estimation from \eqref{eqn:experiment:estimate:pd:direct} for all values of $\alpha$ tested. 
%
Figure~\ref{fig:estimate:delta:11} illustrates the trade-off between improved privacy and utility loss. Increasing utility loss is observed in all cases as the fraction of users with diverse topic interests increases as the ``signal-to-noise'' ratio of coherent interests to random interests decreases. This is observed when either  \eqref{eqn:experiment:estimate:pd:direct} or \eqref{eqn:verify:pd:3a}, for all values of $\alpha$,  are used as estimators.
Using \eqref{eqn:verify:pd:3a} is observed to under-estimate utility loss over all datasets tested. In this case \eqref{eqn:verify:pd:3a} should be taken as a best-case guarantee of utility loss and that the actual utility loss will be higher. We note that the ultimate assessment of utility loss is up to the user - if they do not like the personalised content they receive then they can switch to another proxy agent, or stop using the system entirely.
\subsection{Defending Privacy}
We consider a proactive privacy defence strategy of injecting random queries. Between ``true'' queries a user issues ``noise'' queries to every member of the proxy agent pool \emph{other} than their selected best matching proxy agent about topics \emph{other} than their allocated topic of interest.  This defence is motivated by the observation earlier that increased diversity of topic interests among users is reported to increase plausible deniability. By controlling the level of noise injection we hope to limit the associated utility loss. In practice this kind of injection of obfuscating, uninteresting, ``noise'' queries can be performed in the background by users.
\begin{figure}[!ht]
\vspace{2mm}
  \centering 
  \includegraphics[width=\columnwidth, height=6cm]{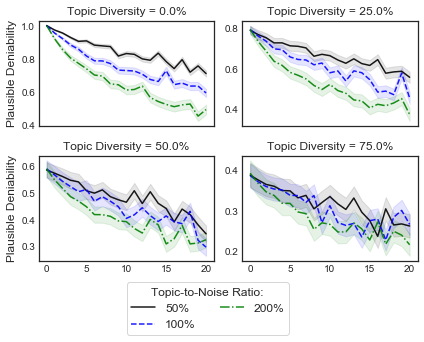}
  \caption{Plausible deniability for different diversity levels in the proxy agent pool for various topic-to-noise ratios. Results are average by topic and over all datasets.\label{fig:defence:pd}}
\end{figure}

\begin{figure}[!ht]
\vspace{2mm}
  \centering 
  \includegraphics[width=\columnwidth, height=6cm]{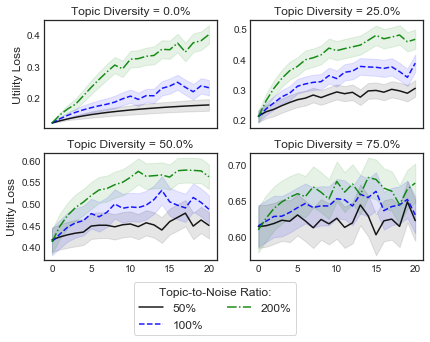}
  \caption{Utility loss for different diversity levels in the proxy agent pool for various topic-to-noise ratios. Results are average by topic and over all datasets.\label{fig:defence:utility:loss}}
\end{figure}

Experimental setups with proxy pools of sizes $3\leq\vert\PP{}\vert\leq30$ and numbers of users $10\leq\vert\UU{}\vert \leq 120$ were configured for each of the test datasets. 
Each proxy agent $p\in\PP{}$ was allocated a topic $c_{p} \in \SET{C}$ as their topic of interest. 
Each user $u\in\UU{}$ was allocated with a target topic of interest $c_u\in\SET{C}$ with setups of $0\%, 25\%, 50\%, 75\%$ and $100\%$ of users having $c_u = c_0$ to model various levels of diversity of topic interests among users.
After a sensitive, true input for topic $c_u$ was issued to a chosen proxy agent, a noise query was constructed where input keywords were drawn at random for topics other than the sensitive user topic $c_u$, and issued to all proxy agents in the pool, except the last chosen proxy agent. To assess the effect of issuing different amounts of noise queries mixed with true queries, ``Topic--to--Noise'' ratios of $50\%, 100\%$ and $200\%$ were also used. So that, for example, in the case of a true-to-noise ratio of $200\%$, $2$ noise queries are issued for every $1$ true queries on average by a user.
Results are reported as averages over $\vert\PP{}\vert$ and $\vert\UU{}\vert$ and topic for compactness and shown for measurements of plausible deniability in Figure~\ref{fig:defence:pd}, and for utility loss in Figure~\ref{fig:defence:utility:loss}. The first plot in each case shows the case when there is $0\%$ diversity of topic interest in the proxy agent pool as a baseline. 

With the random noise injection strategy plausible deniability  against a proxy observer improves steadily during an experimental run for all levels of topic diversity in our experiments.  For all levels of topic diversity, adding more noise results in faster improvement in plausible deniability as expected intuitively. As the topic diversity in the proxy agent pool increases, less random noise is required to produce the same changes in plausible deniability as do larger random noise levels. Intuitively this is to be expected since topic diversity is an indication of the variation in topic interests among users. Standard error in the mean, shown as shaded regions is small, indicating that improved plausible deniability is observed with high confidence for all datasets.   
Utility loss, shown in Figure~\ref{fig:defence:utility:loss}, increases initially and achieves stable levels after $5-10$ input--output steps with the cases where topic diversity is highest reaching a stable level quickest.  Standard error is small in the case of all datasets, suggesting the average values plotted reflect expected behaviour with high confidence.
 
The plausible deniability and utility loss results for $0\%$ topic diversity are a worst-case. Even in this case the utility loss at levels of random noise up to $100\%$ the utility loss is $20\%$ after 20 steps - compared with an improvement in plausible deniability from $100\%$ to $60\%$ on average. As topic diversity increases the improvements in plausible deniability are larger than the associated utility losses in all cases. Taken overall, our results suggest that the benefits to privacy of adopting a strategy of random noise injection outweigh the associated utility losses, with the greatest benefits occurring when the privacy risk from low topic diversity is highest. Run as a background task, injecting random noise by all users in a controlled manner provides a mechanism for enforcing effective topic diversity in the proxy agent pool with corresponding benefits for privacy.
\subsection{Discussion}
The results of the random proxy injection defence in our experiments suggest that once a user is alert to diversity, the 3PS setup can provide balance of probability plausible deniability of topic interests. The method of choosing revealing keyword pairs outlined in Section~\ref{sec:testing:plausible:deniability} provides a practical bound on plausible deniability and is straightforward to apply in practice. In a production setting a browser plug-in could automatically suggest new keywords for inclusion by the user in local keyword dictionary extensions.

To apply \eqref{eqn:define:plausible:deniability:0} in practice, a user also needs a way of confirming that proxy agents are being truthful about the probability estimators it publishes. The notion of \emph{probe queries} was introduced in \cite{mac.P1} to allow a user to test the behaviour of black-box systems without revealing sensitive interests. By checking input--output interactions users can label the observation as sensitive or not and adjust their view of revealing keywords. The techniques introduced in  \cite{mac.P1} can be used to to check for observations that vary from that expected from \eqref{eqn:verify:pd:3a} indicating possible concerns with the estimators distributed by that proxy agent.

Choosing \THA{} to estimate plausible deniability requires care. 
From \eqref{eqn:verify:pd:3a} it follows that 
\begin{align*}
\PHAT_{\alpha}(z\in\SEQ{Z}^{u,c}_{k} \vert z\in\SEQ{Z}_{att,k})
<
\PHAT_{\beta}(z\in\SEQ{Z}^{u,c}_{k} \vert z\in\SEQ{Z}_{att,k})	
\end{align*}%
when $0<\alpha < \beta\leq1$. Choosing $\alpha=1$ to include as many keywords as possible in $\Theta^{u,c}_{u,k}$ is the safest threat detection strategy in our setup here.  We have assumed here that there is no incentive for dishonesty  neither is there any malicious poisoning nor accidental corruption in our setup. In a real-life, production setup when \DICT{X} or \DICT{Y} are partially complete, poisoned or deliberately censored, a user may choose any input--output keywords for \THA{}. We note that the techniques introduced in \cite{mac.P1, mac.P2} provide tools to test when input--output keywords indicate privacy concerns that could be adapted to assist a user with constructing \THA{}.

While our experiments suggest that 3PS can provide acceptable levels of plausible deniability with low utility loss, our results also emphasise the importance of maintaining adequate vigilance to prevent interests in sensitive topics from leaking and taking care to avoid overly revealing content that might compromise plausible deniability when user interests are known.
%
%
\section{Related Work}
\label{sec:related}
The potential for privacy concerns in recommender systems are well known in the literature. For example, Shilling attacks are discussed in \cite{Lam:2004:SRS:988672.988726}; Sybil attacks to determine user preferences in \cite{Calandrino:2011:YML:2006077.2006768}; Shilling attacks to sabotage recommendations in \cite{Lam:2006:YTY:2094770.2094773}, using auxiliary information to de-anonymise Netflix data \cite{DBLP:journals/corr/abs-cs-0610105} and references therein.

Privacy preserving techniques in recommendation systems have largely focused on how to incorporate privacy into the recommendation process. In \cite{Batmaz:2016:RPF:3010164.3010593}, random perturbation of data is used to develop privacy-preserving frameworks for collaborative filtering methods. In \cite{Boutet:2016:PDC:2975427.2975447}, profile obfuscation together with a randomised dissemination protocol are employed. Another approach is to distribute the recommendation process by including a trusted intermediate agent between user and backend system, such as \cite{aimeur2008lambic}. In \cite{McSherry:2009:DPR:1557019.1557090}, differential privacy is incorporated into the algorithms used in the Netflix prize competition to produce privacy preserving recommendations.      

Grouping users behind intermediate layers is a well studied privacy technique. Protecting the sensitivity of user data, and particularly of user profiles exposed to the online system, by grouping users behind a proxy layer is defined as \emph{Level 2 Privacy} in the classification scheme of online privacy approaches in \cite{shen2007privacy}.  In \cite{petit2014towards} a third-party, privacy-proxy hosts a group profile  where the privacy-proxy performs aggregation over multiple user activities to produce the group profile. In \cite{shou2014supporting} profile generalisation is achieved locally on a user's machine via a user-side privacy-proxy layer where the group profile is learned from a globally accessible taxonomy of topics. Obfuscating user data through profile generalisation is studied extensively in the literature. Approaches typically obfuscate or mask user interactions with search engines with the aim of disrupting online profiling and personalisation. PEAS, \cite{Petit:2014:TEA, PEAS}, combines local obfuscation with a privacy-proxy to provide unlink-ability between user and query. Shuffling user profiles as a counter to unwanted profiling in \cite{Biega:2017:PTS:3077136.3080830}. The approach  introduces a trusted third party server to shuffle individual profiles among a pool of users without regard for protecting utility. A common consideration for generalisation approaches in these works is how to provide minimally sufficient common structure to effectively generalise user interests without incurring unacceptable loss of utility. This is commonly solved by distributing generalised usage data, allowing users to create statistical patterns of usage. For example, in \cite{shou2014supporting}, a global dictionary is used that includes statistics on frequency of occurrence of concepts in it. In \cite{petit2014towards}, an intermediate privacy-proxy server distributes similar usage statistics allowing users to construct statistical models of usage.

There are examples of website proxies offering privacy preserving services to access mainstream search engines on the Internet. Two of the better known are DuckDuckGo hosted in the US on Amazon Web Services, \cite{inc.duckduckgo_duckduckgo_2018}, and StartPage hosted privately in the Netherlands, \cite{holding_bv_surfboard_startpage_2018}. Functionally both are similar, encrypting traffic via https, and employing POST and re-direct techniques to obfuscate requests. Both claim to relieve so-called filter-bubbles, \cite{Pariser:2011:FBI:2029079}, by aggregating results from several source systems. In both DuckDuckGo and StartPage the proxy user profile adopted by users of both systems is global. Personalised content such as advertising that is displayed on search result pages is correspondingly generic. 

Protecting users from individual re-identification often combines encryption, hashing and noise addition on the local user machine. Common challenges in designing privacy protection at individual user level is that they can be computationally prohibitive and require substantial user management for locally maintained dictionaries of queries, features or URLS accessed by the user. In \cite{mor2015bloom} user interests with added noise are locally encoded via a Bloom filter instead of in a traditional cookie. Obfuscation through noise is used in \cite{Nikiforakis:2013} where fake URLs drawn from a user-specified dictionary are injected into the user history to confuse an adversary. GooPIR, \cite{domingo2009h,SaNchez:2013:KSC:2383079.2383150}, attempts to disguise a user's ``true''  queries by adding masking features directly into a true query before submitting to a recommender system. Results are filtered to extract items that are relevant to the user's original true query.  Query obfuscation and masking is addressed in \cite{Ahmad:2016} user queries are hidden within a stream of at least $k$ `cover queries' to provide a form of k-anonymity. PWS, \cite{balsa2012ob}, and TrackMeNot, \cite{howe2009trackmenot,peddinti2010privacy}, inject distinct noise queries into the stream of true user queries during a user query session, seeking to achieve acceptable privacy while not overly compromising overall utility. 

There is evidence that users are concerned about their privacy on the Web but do not always reflect this concern in their online behaviours, \cite{acquisti2015privacy}. In \cite{DBLP:conf/imc/PujolHF15},  in-the-wild measurements of user interactions with Ad blocking technologies suggest that users overwhelmingly accept default settings and do not install updates such as whitelists. The conclusion reached is that technologies for user privacy must be effective, but also unobtrusive and simple to maintain. By comparison with users, online systems have proven alert and adaptable in responding to attempts to protect privacy at individual user level. Stateful (cookie) and stateless (fingerprinting) tracking is widespread on the web. In \cite{Bielova:2017:WTT:3133956.3136067, binns2018measuring, englehardt2016online, narayanan2017princeton} separate studies of 1 million websites reveal widespread data exchange among third parties, stateful tracking from third-party cookie spawning and stateless fingerprint-based tracking. In \cite{binns2018measuring} users are observed  to be tracked by multiple entities in tandem on the web.

Search engine algorithm evolution is a continuous ``arms-race'', as evidenced in the case of Google, for example, by major algorithm changes such as \emph{Caffeine} and \emph{Search+ Your World} included additional sources of background knowledge from Social Media, improved filtering of content such as \emph{Panda} to counter spam and content manipulation. More recently semantic search capability has been added through \emph{Knowledge Graph} and \emph{HummingBird}, \cite{search_timeline_1}, \cite{search_history}, \cite{search_history_2}. The importance of personalising content in the online arms-race is further underlined by the continuing arms-race between Ad-blockers and web-site owners. Anti-Ad-blockers are discussed in \cite{nithyanand2016adblocking, mughees2017detecting} in a study of 100,000 popular websites finding evidence that web-site owners are making visible changes to content in when Ad-blockers are detected. In a small number of cases pop-ups are presented that cannot be dismissed until the Adblocking software is disabled. 
\section{Discussion and Conclusions}
\subsection{Discussion}
\label{sec:assess:pool:sz}
Accessing online systems via shared proxies in 3PS appears to provide a natural form of ``hiding on the crowd'' privacy once there is sufficient diversity of users and input--output interactions.  Hence, when the 3PS architecture is used, the main requirement to obtain privacy is to ensure sufficient diversity. Quantities here are expressed in terms of probabilities, with randomness in the process of observing input--output interactions arising from randomness in how the user chooses inputs and any randomness in the system response. This means that practical estimation of these probabilities requires a model of user inputs and system outputs, perhaps derived from observed behaviour but in any case introducing a degree of risk that the model is inadequate and the calculated probability values inaccurate. 

Our experiments indicate that the need to maintain a level of engagement and alertness with respect to individual online privacy is an unavoidable feature of online existence. A framework such as 3PS provides tools to help an engaged user to detect unwanted effects such as affinity in the proxy agent pool but the decision to engage and to take action is an unavoidably personal responsibility. As already discussed, personal judgements regarding risk seem intrinsic to discussions of privacy.  
\subsection{Conclusions}
Through 3PS we provide a user with the capability to achieve anonymity by adopting group identities. We provide a method to decide on the optimal choice of group identity from a pool of proxy identities. The method we develop is both efficient and scalable, and does not require a user to reveal information about their interest in sensitive topics. 

We define a threat model based on notions of increasingly powerful observers with access to various levels of information about the 3PS system. Using the associated attack models we show that 3PS offers users a high degree of protection for their interests in sensitive topics. 

Mass personal data collection is a persistent feature of online systems that has been justified as required for personalisation. Through the framework of the 3PS system our results suggest that much less personal data collection is required for adequate personalisation. This has significant implications for online providers in light of legislation such as GDPR that requires data to be limited to that which is proportionate to the purpose of collection.

Our experiments show that once diversity of likely interests is maintained across  the proxy pool, 3PS provides high levels of protection for users while providing satisfactory personalisation. The 3PS framework provides readily implementable techniques to decide whether a particular choice of proxy agent is overly revealing of likely interest in a topic of their choice so that lack of diversity in the proxy pool is detectable by users without requiring additional infrastructure such as trusted intermediate parties. The defence of injecting noise queries is observed to improve plausible deniability while maintaining levels of utility. Automation of noise injection as well as techniques such as automated suggestion of new keywords through browser plug-in capabilities mean that 3PS can be implemented with relatively little intrusiveness on the user side by automating through, for example, a browser plug-in.

The fast-convergence and high accuracy of the proxy agent selection method, observed in our experiments indicate that 3PS can provide a safe and scalable solution that requires little retro-fitting to work with existing systems.

Overall, our results indicate 3PS is a promising first step and more research is required in large-scale production environments. Directions for future research include, undertaking a practical program of applied research to scale 3PS to a production implementation, investigating how non-text inputs and outputs such as image can be accommodated in 3PS and incorporating 3PS into a larger framework of practical privacy tools to provide robust end-to-end protection for users.  

%
\bibliographystyle{unsrt}
\begin{scriptsize}
\bibliography{biblio}
\end{scriptsize}
\end{document}